\documentstyle[12pt,epsf]{article}
\newcommand{\bfig}{\begin{figure}}
\newcommand{\efig}{\end{figure}}

\newcommand{\maru}{\mbox{\tiny$\stackrel{\circ}{\scriptstyle\circ}$}}

\newcommand{\pdot}{{\displaystyle{\raisebox{-1.5ex}[0.25ex]{$\cdot$}
     \atop\raisebox{0.6ex}[0.25ex]{$\scriptstyle (p)$}}}}
\newcommand{\prdot}{{\displaystyle{\raisebox{-1.5ex}[0.25ex]{$\cdot$}
     \atop\raisebox{0.6ex}[0.25ex]{$\scriptstyle (p^{\prime})$}}}}
\newcommand{\ppdot}{{\displaystyle{\raisebox{-1.5ex}[0.25ex]{$\cdot$}
     \atop\raisebox{0.6ex}[0.25ex]{$\scriptstyle (p,p^{\prime})$}}}}

\newcommand{\kuroten}{{\displaystyle{\raisebox{-1.5ex}{$\odot$}
     \atop\raisebox{0.3ex}[1.25ex]{$\scriptstyle(p,p^{\prime})$}}}}

\def\ap{\alpha^{\prime}}

\def\12{\frac{1}{2}}
\def\bea{\begin{eqnarray}}
\def\eea{\end{eqnarray}}
\def\ba{\begin{array}}
\def\ea{\end{array}}

\def\one-loop{\mbox{\scriptsize one-loop}}

\def\G{\Gamma}

\catcode`\@=11

\@addtoreset{equation}{section}
\def\theequation{\arabic{section}.\arabic{equation}}

\def\@normalsize{\@setsize\normalsize{15pt}\xiipt\@xiipt
\abovedisplayskip 14pt plus3pt minus3pt%
\belowdisplayskip \abovedisplayskip
\abovedisplayshortskip  \z@ plus3pt%
\belowdisplayshortskip  7pt plus3.5pt minus0pt}

\def\small{\@setsize\small{13.6pt}\xipt\@xipt
\abovedisplayskip 13pt plus3pt minus3pt%
\belowdisplayskip \abovedisplayskip
\abovedisplayshortskip  \z@ plus3pt%
\belowdisplayshortskip  7pt plus3.5pt minus0pt
\def\@listi{\parsep 4.5pt plus 2pt minus 1pt
            \itemsep \parsep
            \topsep 9pt plus 3pt minus 3pt}}

\def\underline#1{\relax\ifmmode\@@underline#1\else
        $\@@underline{\hbox{#1}}$\relax\fi}
\@twosidetrue

\relax

\catcode`@=12

\catcode`\@=11

\def\section{\@startsection{section}{1}{\z@}{3.5ex plus 1ex minus
   .2ex}{2.3ex plus .2ex}{\large\bf}}

\def\thesection{\Roman{section}.}

\def\appendix{\setcounter{section}{0}
        \def\thesection{Appendix }
        \def\theequation{\Alph{section}.\arabic{equation}}}

\def\ps@headings{\def\@oddfoot{}\def\@evenfoot{}
\def\@oddhead{\hbox{}\hfill
        \makebox[.5\textwidth]{\raggedright\ignorespaces --\thepage{}--
        \hfill {}}}
\def\@oddhead{\hbox{}\hfill --\thepage{}-- \hfill
        {}}
\def\@evenhead{\@oddhead}
\def\subsectionmark##1{\markboth{##1}{}}
}

\ps@headings

\catcode`\@=12

\relax

\def\figcap{\section*{Figure Captions\markboth
        {FIGURECAPTIONS}{FIGURECAPTIONS}}\list
        {Fig. \arabic{enumi}:\hfill}{\settowidth\labelwidth{Fig. 999:}
        \leftmargin\labelwidth
        \advance\leftmargin\labelsep\usecounter{enumi}}}
 \relax
\def\tablecap{\section*{Table Captions\markboth
        {TABLECAPTIONS}{TABLECAPTIONS}}\list
        {Table \arabic{enumi}:\hfill}{\settowidth\labelwidth{Table 999:}
        \leftmargin\labelwidth
        \advance\leftmargin\labelsep\usecounter{enumi}}}
 \relax
\def\reflist{\section*{References\markboth
        {REFLIST}{REFLIST}}\list
        {[\arabic{enumi}]\hfill}{\settowidth\labelwidth{[999]}
        \leftmargin\labelwidth
        \advance\leftmargin\labelsep\usecounter{enumi}}}
 \relax

\catcode`\@=11

\def\ps@headings{\def\@oddfoot{}\def\@evenfoot{}
\def\@oddhead{\hbox{}\hfill
        \makebox[.5\textwidth]{\raggedright\ignorespaces --\thepage{}--
        \hfill {}}}
\def\@evenhead{\@oddhead}
\def\subsectionmark##1{\markboth{##1}{}}
}

\ps@headings

\relax

\newskip\humongous \humongous=0pt plus 1000pt minus 1000pt

\newif\ifdtup

\def\beq{\begin{equation}}
\def\eeq{\end{equation}}

\def\beqn{\begin{eqnarray}}
\def\eeqn{\end{eqnarray}}
\relax

\def\G2{{\; \rm GeV/}c^2}
\def\G{\; \rm GeV}

\def\dotx{\dotx{\dot\overline{x}}}

\relax

\begin{document}

%
%
\begin{titlepage}

\renewcommand{\thefootnote}{\fnsymbol{footnote}}

\begin{flushright}
      \normalsize
      November, 2000  \\     
     OU-HET 368, \\ 
         hep-th/0011028  \\
\end{flushright}

%
\begin{center}
  {\large\bf Note on  Open String/D-brane System  \\ 
  and Noncommutative Soliton }%
\footnote{This work is supported in part
 by the Grant-in-Aid  for Scientific Research
(12640272, 12014210) from the Ministry of Education,
Science and Culture, Japan}
\end{center}

\vfill

\begin{center}
    { { H. Itoyama}\footnote{e-mail address:
                               itoyama@funpth.phys.sci.osaka-u.ac.jp} }\\
\end{center}

\vfill

\begin{center}
      \it  Department of Physics,
        Graduate School of Science, Osaka University,\\
        Toyonaka, Osaka 560-0043, Japan
\end{center}

\vfill


\begin{abstract}
   This is a summary of a series of papers \cite{CIMM,CIMM2,CIMM3}
 written with B. Chen, T. Matsuo and 
 K. Murakami on a $p-p^{\prime}, (p<p^{\prime})$ open string
 with $B_{ij}$ field, which
 has led us to the explicit identification of the $Dp$-brane with the
 noncommutative projector soliton via the gaussian damping factor.
    A lecture given at Summer Institute 2000,  FujiYoshida, Yamanashi, Japan, 
 at August 7-14, 2000.

\end{abstract}

\vfill

\setcounter{footnote}{0}
\renewcommand{\thefootnote}{\arabic{footnote}}

\end{titlepage}


\section{Introduction}

The $p-p^{\prime},$ $(p<p^{\prime})$ open string is
 one of our favorite systems to discuss gauge fields and topological defects
  appearing on the worldvolume of the bigger brane. The presence of $B$
 field
 makes
 this situation more interesting and, in some sense, more tractable.
 ( See, for example, \cite{Callan1, SW}.) 
  A little less than a year ago, we have started to look at this system
 in
 string perturbation
  theory. I would like to present several interesting results coming
 out from our computation
  and understading of the phenomena.

  There are two distinct points of view to the system.
\begin{enumerate}
\item    the view from the smaller ($Dp$) brane  : 
  this may be called moduli space point of view.
\item     the view from the bigger ($Dp^{\prime}$) brane  :
       this may be called spacetime point of view, which
 we have put forward in our investigation.
\end{enumerate}
In what follows, the D$p$-brane extends in
the $(x^{0},x^{1},\ldots,x^{p})$-directions and the D$p^{\prime}$-brane
extends in the $(x^{0},x^{1},\ldots,x^{p^{\prime}})$-directions
with the D$p$-brane inside.
The D$p$-brane worldvolume contains the boundary $\sigma=0$ 
while the D$p^{\prime}$-brane
worldvolume contains the boundary $\sigma=\pi$.

\section{Description of the System and Worldsheet Properties}

\subsubsection{the action of the NSR superstring in the constant $B$
 background:}
\begin{equation}
 S=\frac{1}{2\pi}\int d^{2}{\xi}
   \int d\theta d\overline{\theta} \left(
     g_{\mu\nu}+2\pi\alpha^{\prime}B_{\mu\nu}
   \right)\overline{D} {\bf X}^{\mu}({\bf z},\overline{\bf z})
           D {\bf X}^{\nu}({\bf z},\overline{\bf z})~,
\label{eq:action-1}
\end{equation}
where ${\bf z}=(z,\theta)$ and
$\overline{\bf z}=(\overline{z},\overline{\theta})$,
$ D=\frac{\partial}{\partial \theta}
+ \theta \frac{\partial}{\partial z}$ and
$ \overline{D}=\frac{\partial}{\partial \overline{\theta}}
  +\overline{\theta}\frac{\partial}{\partial \overline{z}}$,
$z=\xi^{1}+i\xi^{2}$ and
$\overline{z}=\xi^{1}-i\xi^{2}$ and
 $z=e^{\tau+i\sigma}$ and $\overline{z}=e^{\tau-i\sigma}$.
\begin{equation}
 g_{\mu\nu} = \left(
\begin{array}{ccccccc}
 \multicolumn{1}{c|}{-1}& & & & & & \\ \cline{1-4}
 \multicolumn{1}{c|}{ }&\multicolumn{3}{c|}{ }& & &   \\
 \multicolumn{1}{c|}{ }& &{g_{ij}}&\multicolumn{1}{c|}{}& & &  \\
 \multicolumn{1}{c|}{ }& &{ }&\multicolumn{1}{c|}{}& & & \\ \cline{2-7}
     & & &\multicolumn{1}{c|}{ } &1 & & \\
     & & &\multicolumn{1}{c|}{ } & & \ddots & \\
     & & &\multicolumn{1}{c|}{ } & & & 1 
\end{array}
\right)~,
\qquad\quad
g_{ij}=\varepsilon \delta_{ij}\quad(i,j=1,\ldots,p^{\prime})~,
\label{eq:g-munu}
\end{equation}
\begin{equation}
 B_{ij}=\frac{\varepsilon}{2\pi\alpha^{\prime}}\left(
 \begin{array}{ccccc}
  0&b_{1} & & & \\
  {}-b_{1}&0& & & \\
   & & 0 & b_{2}& \\
   & & -b_{2} & 0 & \\
   & & & & \ddots
 \end{array}\right)
\quad (i,j=1,\ldots,p^{\prime})~,
\quad\mbox{otherwise $B_{\mu\nu}=0$}~.  \label{eq:B-ij}
\end{equation}

\subsubsection{the boundary conditions for the string coordinates
 in the NS sector:}
\begin{eqnarray}
&& \left. D{\bf X}^{0}-\overline{D}{\bf X}^{0}
\right|_{\sigma=0,\pi~\theta=\overline{\theta}}=0~,
\quad
\left. D{\bf X}^{p^{\prime}+1,\ldots,9}
       +\overline{D}{\bf X}^{p^{\prime}+1,\ldots,9}
\right|_{\sigma=0,\pi~\theta=\overline{\theta}}=0~,\nonumber\\
&& \left. g_{kl}(D{\bf X}^{l}-\overline{D}{\bf X}^{l})
          +2\pi\alpha^{\prime}B_{kl}
            (D{\bf X}^{l}+\overline{D}{\bf X}^{l})
   \right|_{\sigma=0,\pi~\theta=\overline{\theta}}=0
\quad (k,l=1,\ldots,p)\nonumber\\
&&\left. D{\bf X}^{i}+\overline{D}{\bf X}^{i}
  \right|_{\sigma=0~\theta=\overline{\theta}}
 =\left.g_{ij}(D{\bf X}^{j}-\overline{D}{\bf X}^{j})
  +2\pi \alpha^{\prime}B_{ij}(D{\bf X}^{j}+\overline{D}{\bf X}^{j})
  \right|_{\sigma=\pi~\theta=\overline{\theta}}=0\nonumber\\
&& \hspace{24em}(i,j=p+1,\ldots,p^{\prime})~.
\end{eqnarray}
 One may say that the scale of noncommutativity is contained in
 these boundary conditionss.

\subsubsection{quantization and the mode expansion}
 We concentrate only on the $x^{i}$-directions
$(i=p+1,\ldots,p^{\prime})$.
We complexify the string coordinates
${\bf X}^{i}({\bf z},\overline{\bf z})$ in these directions as
\begin{eqnarray}
 {\bf Z}^{I}({\bf z,\overline{z}})&=&
    {\bf X}^{2I-1}({\bf z,\overline{z}})
    +i{\bf X}^{2I}({\bf z,\overline{z}})
  = \sqrt{\frac{2}{\alpha^{\prime}}}Z^{I}(z,\bar{z})
      +i\theta\Psi^{I}(z)
      +i\overline{\theta}\widetilde{\Psi}^{I}(\overline{z})~,\nonumber\\
\overline{\bf Z}^{\overline{I}}({\bf z,\overline{z}})&=&
     {\bf X}^{2I-1}({\bf z,\overline{z}})
    {}-i{\bf X}^{2I}({\bf z,\overline{z}})
 = \sqrt{\frac{2}{\alpha^{\prime}}}
         \overline{Z}^{\overline{I}}(z,\overline{z})
   +i \theta\overline{\Psi}^{\overline{I}}(z)
   +i \overline{\theta}
         \widetilde{\overline{\Psi}}^{\overline{I}}(\overline{z})~,
\end{eqnarray}
where $I,\overline{I}=\frac{p+2}{2},\ldots,\frac{p^{\prime}}{2}$.
\begin{eqnarray}
 Z^{I}(z,\overline{z})
  =i\sqrt{\frac{\alpha^{\prime}}{2}}\sum_{n\in {\bf Z}}
     \frac{\alpha^{I}_{n-\nu_{I}}}{n-\nu_{I}}
     \left(z^{-(n-\nu_{I})}-\overline{z}^{-(n-\nu_{I})}\right)~,\nonumber\\
 \overline{Z}^{\overline{I}}(z,\overline{z})
   =i\sqrt{\frac{\alpha^{\prime}}{2}}\sum_{m\in {\bf Z}}
     \frac{\overline{\alpha}^{\overline{I}}_{m+\nu_{I}}}{m+\nu_{I}}
     \left( z^{-(m+\nu_{I})}-\overline{z}^{-(m+\nu_{I})} \right)~,
\end{eqnarray}
where $\nu_{I}$ are defined by
\begin{equation}
 e^{2\pi i \nu_{I}}=-\frac{1+ib_{I}}{1-ib_{I}}~,
\quad 0< \nu_{I} < 1~.
\end{equation}
Now we can introduce the open string metric
$G^{IJ}$, $G^{\overline{I}\overline{J}}$, $G^{I\overline{J}}$
and $G^{\overline{I}J}$
concerning the $x^{p+1},\ldots,x^{p^\prime}$ directions,
\begin{equation}
G^{IJ}=G^{\overline{I}\overline{J}}=0~,\quad
G^{I\overline{J}}=G^{\overline{J}I}
  =\frac{2}{\varepsilon(1+b^2_I)}\delta^{I\overline{J}}~.
\end{equation}
The commutation relations are
\begin{eqnarray}
&&[\alpha^{I}_{n-\nu_{I}} \, , \,
        \overline{\alpha}^{\overline{J}}_{m+\nu_{J}}]
   =\frac{2}{\varepsilon}\delta^{I\overline{J}}(n-\nu_{I})
                   \delta_{n+m}~,\nonumber\\
&&\{b^{I}_{r-\nu_{I}}\, , \,\overline{b}^{\overline{J}}_{s+\nu_{J}}\}
     =\frac{2}{\varepsilon} \, \delta^{I\overline{J}}\delta_{r+s}~,
 \qquad
  \{d^{I}_{n-\nu_{I}}\, , \,\overline{d}^{\overline{J}}_{m+\nu_{J}}\}
      =\frac{2}{\varepsilon} \, \delta^{I\overline{J}}\delta_{n+m}~.
\end{eqnarray}

\subsubsection{the oscillator vacuum}
The oscillator vacuum $\left|\sigma, s \right\rangle
 \equiv \left|\sigma \right\rangle \bigotimes
 \left| s \right\rangle$:
\begin{equation}
\left| \sigma \right\rangle 
  = \bigotimes_{I}\left|\sigma_{I}\right\rangle
\qquad
\mbox{with}
\qquad
\left\{
\begin{array}{ll}
 \alpha^{I}_{n-\nu_{I}}\left|\sigma_{I}\right\rangle=0
&n>\nu_{I} \\
 \overline{\alpha}^{\overline{I}}_{m+\nu_{I}}
        \left|\sigma_{I}\right\rangle=0
&m>-\nu_{I}
\end{array}
\right.~. \label{eq:osci-vac-z}
\end{equation}
\begin{equation}
 \left|s\right\rangle = \bigotimes_{I}\left|s_{I}\right\rangle
\qquad
\mbox{with}
\qquad
\left\{
\begin{array}{ll}
 b^{I}_{r-\nu_{I}}\left| s_{I} \right\rangle =0
& r \geq \frac{1}{2} \\
 \overline{b}^{\overline{I}}_{s+\nu_{I}}\left|s_{I}\right\rangle =0
& s \geq \frac{1}{2}
\end{array}
\right.~,\label{eq:osci-vac-ns}
\end{equation}
A twist field $\sigma^{+}_{I}(\xi^{1})$
and an anti-twist field $\sigma^{-}_{I}(\xi^{1})$, both of
which are mutually non-local with respect to
$Z^{I}$ and $\overline{Z}^{I}$,
are located at the origin and at  infinity on the plane
 respectively \cite{DFMS, GNS}.
They create a branch cut between themselves.
The incoming vacuum $\left|\sigma_{I}\right\rangle$
defined in eq.~(\ref{eq:osci-vac-z}) should be interpreted
as being excited from the $SL(2,{\bf R})$-invariant vacuum
$|0\rangle$ by the twist field $\sigma^{+}_{I}$:
\begin{equation}
 \left|\sigma_{I}\right\rangle
 = \lim_{\xi^{1}\rightarrow 0}\sigma^{+}_{I}(\xi^{1})
   \,|0\rangle~.
\end{equation}
 A similar comment applies to the spin field.

\subsubsection{spectrum}
 Reflecting the vacuum sea filling,
  the spectrum of the $p-p^{\prime}$ system cannot be given generically. 
  The spectrum of each individual case of $p-p^{\prime}$
 has been fully analyzed in \cite{CIMM}. ( See also \cite{SW} for
 the $0-4$ case.)
 If we finetune the sign of
 $b_I$,  a large number of light states appear in the limit. To be
 more  precise, these light states are obtained by acting
 the several low-lying fermionic modes on  the oscillator
 vacuum  and  multiplying by an
 arbitrary polynomial consisting of the lowest bosonic mode.
   This latter bosonic mode is the one which has failed
to become a momentum due to the boundary condition of the $p-p^{\prime}$
open string and is responsible for an infinite number of nearly degenerate
 low-lying states.

\subsubsection{two-point function on superspace}
Let
\beq
\mbox{\boldmath$\cal G$}^{I\overline{J}}
   \left({\bf z}_{1},\overline{\bf z}_{1}
         | {\bf z}_{2},\overline{\bf z}_{2}\right)
 \equiv
   \left\langle \sigma,s \right|{\cal R}
     {\bf Z}^{I}({\bf z}_{1},\overline{\bf z}_{1})
      \overline{\bf Z}^{\overline{J}}
({\bf z}_{2},\overline{\bf z}_{2})
  \left| \sigma,s\right\rangle \;\;.
\label{eq:propCIMM-1}
\eeq
When restricted onto the worldsheet boundary on
the D$p^{\prime}$-brane worldvolume,
this becomes
\begin{equation}
 \mbox{\boldmath${\cal G}$}^{I\overline{J}}
 \left(-e^{\tau_{1}},\theta_{1} | -e^{\tau_{2}},\theta_{2}\right)
=4 G^{I\overline{J}}
  \,{\cal H}\left(\nu_{I};
         \frac{e^{\tau_{1}}}{e^{\tau_{2}}-\theta_{1}\theta_{2}}\right)
+\epsilon(\tau_{1}-\tau_{2}) \frac{4}{\varepsilon}
 \frac{\delta^{I\overline{J}}}{1+b_{I}^{2}}\pi b_{I}~,
\label{eq:propCIMM-3}
\end{equation}
where ${\cal H}(\nu;z)$ is defined by using the hypergeometric
series as
\begin{equation}
 {\cal H}(\nu;z)=\left\{
\begin{array}{ll}
 \displaystyle{\cal F} \left(1-\nu_{I};\frac{1}{z}\right)
     {}-\frac{\pi}{2} b_{I}
   =\sum_{n=0}^{\infty}\frac{z^{-n-1+\nu_{I}}}{n+1-\nu_{I}}
     {}-\frac{\pi}{2} b_{I}&
     \mbox{ for $|z| > 1$}\\[3ex]
 \displaystyle{\cal F} \left( \nu_{I};z \right)+\frac{\pi}{2}b_{I}
  =\sum_{n=0}^{\infty}\frac{z^{n+\nu_{I}}}{n+\nu_{I}}
   +\frac{\pi}{2}b_{I} &
     \mbox{ for $|z| <1$} \;\;.
\end{array}
\right.   \label{eq:h-prop}
\end{equation}
The function ${\cal F}(\nu\,;\,z)$ is defined as
\begin{equation}
 {\cal F}(\nu\,;\, z) =\frac{z^{\nu}}{\nu}\, F(1,\nu;1+\nu;z)
 =\sum_{n=0}^{\infty}\frac{1}{n+\nu}z^{n+\nu}~,
\end{equation}
and $F(a,b;c;z)$ is the hypergeometric function.
By using the hypergeometric series, we find that
\begin{equation}
 {\cal F}(1-\nu_{I};1)-{\cal F}(\nu_{I};1)
=- \sum_{n=-\infty}^{\infty}\frac{1}{n+\nu_{I}}
=-\pi \cot \left( \pi\nu_{I} \right)=\pi b_{I}~.
\label{eq:hyperseries}
\end{equation}
This means that, as is pointed out in \cite{CIMM},
the noncommutativity on the D-brane worldvolume in the $p$-$p^{\prime}$
system is the same as that in the $p$-$p$ system.

\subsubsection{renormal ordering and subtracted two-point function}
 The contents of this subsection are crucial ingredients of the derivation of
 the string amplitudes presented in next section and are responsible for
 the major spacetime properties of the system. 
We have two types of vacuum:
the one is the $SL(2,{\bf R})$-invariant vacuum
and the other is the oscillator vacuum.
We will use the symbols $:\ :$ and $\maru~\maru$
to denote the normal orderings
with respect to the $SL(2,{\bf R})$-invariant vacuum
and the oscillator vacuum respectively.
The $\maru~\maru$-normal ordered product
for the free fields in the $x^{i}$-directions
$(i=p+1,\ldots,p^{\prime})$ is
\begin{equation}
 \maru~{\bf Z}^{I}({\bf z}_{1},\overline{\bf z}_{1})
    \overline{\bf Z}^{\overline{J}}({\bf z}_{2},\overline{\bf z}_{2})~
 \maru
   ={\cal R}{\bf Z}^{I}({\bf z}_{1},\overline{\bf z}_{1})
    \overline{\bf Z}^{\overline{J}}({\bf z}_{2},\overline{\bf z}_{2})
   {}-\mbox{\boldmath${\cal G}$}^{I\overline{J}}
           ({\bf z}_{1},\overline{\bf z}_{1}|
            {\bf z}_{2},\overline{\bf z}_{2})~,
\end{equation}
for $I,\overline{J}=\frac{p+2}{2},\ldots,\frac{p^{\prime}}{2}$.
Here $\mbox{\boldmath${\cal G}$}^{I\overline{J}}
           ({\bf z}_{1},\overline{\bf z}_{1}|
            {\bf z}_{2},\overline{\bf z}_{2})$
is the two-point function defined in eq.~(\ref{eq:propCIMM-1}).
The formula of the renormal ordering takes the form of
\begin{equation}
 :\mbox{\boldmath$\cal O$}:
  =\exp\left(\int d^{2}{\bf z_{1}}
           d^{2}{\bf z}_{2} \,
           {\mbox{\boldmath ${\cal G}$}_{\rm sub}}^{I\overline{J}}
                  ({\bf z}_{1},\overline{\bf z}_{1}|
                   {\bf z}_{2},\overline{\bf z}_{2})
   \frac{\delta}{\delta{\bf Z}^{I}({\bf z}_{1},\overline{\bf z}_{1})}
   \frac{\delta}{\delta \overline{\bf Z}^{\overline{J}}
                   ({\bf z}_{2},\overline{\bf z}_{2})}
   \right)
  \maru~\mbox{\boldmath $\cal O$}~\maru~.
\label{eq:reorder}
\end{equation}
Here 
${\mbox{\boldmath$\cal G$}_{\rm sub}}^{I\overline{J}}
    ({\bf z}_{1},\overline{\bf z}_{1}|{\bf z}_{2},\overline{\bf z}_{2})$
is the subtracted two-point function defined as
\begin{eqnarray}
&& {\mbox{\boldmath$\cal G$}_{\rm sub}}^{I\overline{J}}
  ({\bf z}_{1},\overline{\bf z}_{1}|{\bf z}_{2},\overline{\bf z}_{2})
\equiv
  \langle \sigma,s|:{\bf Z}^{I}({\bf z}_{1},\overline{\bf z}_{1})
    \overline{\bf Z}^{\overline{J}}({\bf z}_{2},\overline{\bf z}_{2}):
  |\sigma,s\rangle     \nonumber\\
&& \hspace{1em}=  \mbox{\boldmath$\cal G$}^{I\overline{J}}
     ({\bf z}_{1},\overline{\bf z}_{1}|{\bf z}_{2},\overline{\bf z}_{2})
  {}-{\bf G}^{I\overline{J}}
     ({\bf z}_{1},\overline{\bf z}_{1}|{\bf z}_{2},\overline{\bf z}_{2})~,
\label{eq:G-sub}
\end{eqnarray}
  ${\bf G}^{I\overline{J}} ({\bf z}_{1},
\overline{\bf z}_{1}|{\bf z}_{2},\overline{\bf z}_{2})$
is the two-point function defined with respect to the
 $SL(2,{\bf R})$-invariant vacuum  and take the well-known form
 \cite{Callan1,SW}.

\section{Scattering Amplitudes}
 Let us specify the process of our interest.
At an initial state, we prepare a $Dp$-brane which is at rest and which lies
in the worldvolume of a $Dp^{\prime}$-brane.
  We place the tachyon (the lowest mode) of a $p-p^{\prime}$ open string
 which carries a
 momentum $k_{1\mu},$ $\mu = 0\cdots p$
along the $Dp$-brane worldvolume. In addition, $N-2$ noncommutative 
$U(1)$ photons carrying momenta $k_{a M}$ 
$a=3\cdots N,\; M= 0 \cdots p^{\prime}$ in $p^{\prime} +1$ dimensions are 
present.  They get absorbed into the $Dp$-brane. At a final state, the
 $Dp$-brane
is found to be present and the momentum of the tachyon is measured to be
 $-k_{2\mu}$ along the $Dp$-brane worldvolume.  We will examine the tree
 scattering amplitude of this process both from string perturbation theory of
 the D-brane/open string system in the zero slope limit
 and from perturbation theory of the field theory action proposed
 in \cite{CIMM2}.  We will find that computations from both sides in fact
 agree by identifying the $Dp$-brane with an initial/final configuration
  representing a noncommutative soliton.   
Let
\begin{eqnarray}
\kappa_{I} = \frac{1}{2}\left(k_{2I-1}-ik_{2I}\right)~,
&&
\overline{\kappa}_{\overline{I}}
 =\frac{1}{2}\left(k_{2I-1}+ik_{2I}\right)~;\nonumber\\
e_{I}(k)=\frac{1}{2}\Big(\zeta_{2I-1}(k)-i\zeta_{2I}(k)\Big)~,
&&
\overline{e}_{\overline{I}}(k)
  =\frac{1}{2}\Big(\zeta_{2I-1}(k)+i\zeta_{2I}(k)\Big)~.
\end{eqnarray}
Let $({\rm NC})$ denote
\begin{eqnarray}
({\rm NC}) &=&\sum_{1\leq a<a^{\prime}\leq N}\frac{i}{2}
              \epsilon(x_{a}-x_{a^{\prime}})
          \sum_{i,j=1}^{p}\theta^{ij}k_{ai}k_{a^{\prime}j}
\nonumber\\
 &&-\sum_{3\leq c<c^{\prime} \leq N}
   \epsilon(x_{c}-x_{c^{\prime}})
   \sum_{I,\overline{J}}\alpha^{\prime}
    \frac{2\delta^{I\overline{J}}\pi b_{I}}{\varepsilon(1+b_{I}^{2})}
    \left(\kappa_{cI}\overline{\kappa}_{c^{\prime}\overline{J}}
          {}-\overline{\kappa}_{c\overline{J}}\kappa_{c^{\prime}I}
    \right)\nonumber\\
 &=&\sum_{1\leq a<a^{\prime}\leq N}\frac{i}{2}
      \epsilon (x_{a}-x_{a^{\prime}})\sum_{\mu,\lambda=0}^{p^{\prime}}
      \theta^{\mu\lambda}k_{a\mu}k_{a^{\prime}\lambda}~,
\label{eq:realnc}
\end{eqnarray}
with $k_{1j}=k_{2j}=0$ for $(j=p+1,\ldots,p^{\prime})$.
Space permits us only the final form of the $N$ point amplitude:
\begin{eqnarray}
\label{eq:finalform}
\lefteqn{
A_N = c (2\pi)^{p+1}\prod^p_{\mu=0}
\delta\left(\sum^N_{e=1}k_{e\mu}\right) \int \prod^N_{a=4}dx_a
\prod^N_{a^\prime=3}d\theta_{a^\prime}d\eta_{a^\prime}
\exp{\cal C}_{a^\prime}(\nu_I)} \nonumber\\
& &\times \prod^N_{c=4}\left[
          x_c^{-\alpha^\prime s_c+\alpha^{\prime} m^2_T}
          (1-x_c)^{2\alpha^{\prime} k_3\pdot k_c}\right]
    \prod_{4\leq c<c^\prime\leq N}
       (x_c-x_{c^\prime})^{2\alpha^{\prime} k_{c^\prime}\pdot k_c}
\nonumber \\
& &\times \prod_{3\leq c<c^\prime\leq N}
   \exp\left[-2\alpha^{\prime}\sum_{I,\overline{J}}G^{I\overline{J}}
     \left\{ \kappa_{cI}\overline{\kappa}_{c^\prime\overline{J}}
            {\cal H}\left(\nu_I; \frac{x_c}{x_{c^\prime}}\right)
           +\overline{\kappa}_{c\overline{J}} \kappa_{c^{\prime}I}
          {\cal H}\left(\nu_{I};\frac{x_{c^{\prime}}}{x_{c}}\right)
     \right\} \right]
\nonumber\\
& &\times \exp\left({\rm NC}\right)
 \exp\Big([0,2]+[2,0]+[1,1]+[2,2]\Big)
\Bigg|_{x_{1}=0,x_{2}=\infty,x_3=1}~.
\end{eqnarray}
This expression is  regarded as an $SL(2,{\bf R})$ invariant integral
(Koba-Nielsen) representation for the amplitude of our concern. 
Let us list several features which are distinct from the corresponding
formula in the case of a $p$-$p$ open string. (See \cite{IM}).
\begin{enumerate}
\item The term denoted by  $\exp\left({\rm NC}\right)$ 
      which originated from the noncommutativity of the worldvolume
      extends into both the $x^{1},\ldots, x^{p}$ directions
      and the remaining $x^{p+1},\ldots,x^{p^\prime}$ directions.
\item To each external vector leg, we have a momentum dependent
      multiplicative factor $\exp {{\cal C}(\nu_I)}$.
\item A new tensor $J$ has appeared.
\item There are parts in the expression
      which are expressible in terms of the momenta of the tachyons, 
      the momenta and the polarization tensors of the vectors
      and $J$ alone, using the inner product with respect to
      the open string metric.
      These parts come, however, with a host of other parts
      which do not permit such generic description in terms of
      the inner product.
\end{enumerate}
The terms containing $\theta_a$ and $\eta_a$ are classified  by
 the number of $\eta_a$ and by the number of $\theta_a$, which we
 designate respectively
 by the first and by the second entry inside the bracket. These are given as
 $[0,2], [2,0], [1,1]$, and $[2,2]$.
For $N=3$ case, we pick up $\theta_3$ and
$\eta_3$ from $[1,1]$ to saturate the Grassmann integrations.
For $N=4$ case, we pick up terms from
$[2,2]+[0,2][2,0]+\frac{1}{2}[1,1]^2$.

\section{The Zero Slope Limit and Noncommutative
 Soliton}
 We focus on the nontrivial zero slope limit of
 the amplitude.\footnote{In this and the next sections, 
the spacetime index
 $M,N \cdots$ run from $0$ to $p^{\prime}$, $\mu, \nu \cdots$ from
 $0$ to $p$ and   $m,n \cdots$  from $p+1$ to $p^{\prime}$.}
The zero slope limit is defined as
\begin{equation}
\begin{array}{rcl}
\alpha' &\sim& \varepsilon^{1/2} \to 0~,  \\
g &\sim& \varepsilon \to 0~, \\
|b_{I}| &\sim& \varepsilon^{-1/2} \to \infty~.
\end{array} \end{equation} 
 This limit keeps $\alpha' b_{I}$ finite:
\beq
\alpha' b_{I} \to \beta_{I}\;\;\;.
\label{eq:psib}
\eeq
  In terms of the open string metric and the noncommutativity parameter,  
this implies
\beq
      \frac{1}{2\pi} \left( J G \theta \right)_{2I-1}^{\;\;2I-1} =
      \frac{1}{2\pi} \left( J G \theta \right)_{2I}^{\;\;2I}  = \beta_{I} \;\;.
\eeq
 In addition, the following limit is taken without loss of generality:
\beq
\label{eq:signb}
\nu\equiv\nu_{\frac{p+2}{2}}  \rightarrow  1  \;, \;\;\;\;
\nu_{\widetilde{I}}  \rightarrow 0  \;, \;\;    \rm{for} \;\;
 \widetilde{I}\neq \frac{p+2}{2}\;\;,
\eeq
 so that
\beq
\label{eq:signb2}
 b_{\frac{p+2}{2}} \to  + \infty \;\;, \;\;\;
 b_{\widetilde{I}} \to  - \infty \;\;.
\eeq

The exponential multiplicative factor $\exp {\cal C}(\left\{  \nu_I \right\})$
 becomes in the zero slope limit
\begin{equation}
  \exp{\cal C}( \left\{  \nu_I \right\} )\to  \exp \left(
-\pi\sum_{I,\bar{J}}\left|\beta_{I}\right|
\kappa_{I}\overline{\kappa}_{\overline{J}}G^{I\overline{J}}  \right)
=  \exp   \left( -\frac{\pi}{2}\sum_{I}\left|\beta_{I}\right|
  \left(k\kuroten k\right)_{I}  \right)
  \equiv D \left(k_{m} \right)
\;\;. \label{eq:damping2}
\end{equation}
This factor is originally associated with each vector
 propagating into the 
$x^{p+1} \sim x^{p^{\prime}}$ directions. We will refer to this as
 gaussian damping factor (g.d.f.).
 Replacing $c \sqrt{\frac{\alpha^{\prime}}{2}}$ by the initial and
 final wave functions of the tachyon and the noncommutative $U(1)$
 photon,
 which we should insert together with the vertex operators,
 we obtain
\beqn
  \lim  {\cal A}_{3}=  (2\pi)^{p+1}\prod_{\mu=0}^{p}
  \delta\left(\sum^3_{a=1}k_{a\mu}\right) 
 \frac{1}{(2\pi)^{p/2}}\frac{1}{\sqrt{2\omega_{{\vec k}_{2}}}}
\frac{1}{(2\pi)^{p/2}}\frac{1}{\sqrt{2\omega_{{\vec k}_{1}}}}
\frac{1}{(2\pi)^{p^\prime/2}}
\frac{1}{\sqrt{2|{\vec k}_{3}|}}   \;\; \nonumber \\
  \times \left\{(k_2-k_1)\pdot \zeta_3-ik_3\ppdot J \zeta_3
\right\}  D(k_{3m})
  e^{\frac{i}{2}\theta^{ij}k_{1i}k_{2j}} \;\;.
\label{eq:lima3}
\eeqn
Let us consider the following factor seen as an exponent of
eq. (\ref{eq:finalform}):
\beqn
\label{eq:factor}
  P_{c^{\prime} c} \equiv -2\alpha^{\prime}
\sum_{I,\overline{J}}G^{I\overline{J}}
     \left( \kappa_{cI}\overline{\kappa}_{c^\prime\overline{J}}
            {\cal H}\left(\nu_I; \frac{x_c}{x_{c^\prime}}\right)
           +\overline{\kappa}_{c\overline{J}} \kappa_{c^{\prime}I}
          {\cal H}\left(\nu_{I};\frac{x_{c^{\prime}}}{x_{c}}\right)
 \right) \:\:.
\eeqn
 We see that, in any region contributing to the zero slope limit,
 this factor $P_{c^{\prime} c}$ contains precisely the identical
  constant piece 
$\displaystyle{-\pi \sum_I|\beta_I|(k_c\kuroten k_{c^\prime})_I }$ in the
 limit.   Multiplying   
$\displaystyle{ \prod_{3\leq c<c^\prime\leq N} \exp \left[
 -\pi \sum_I \mid  \beta_I \mid (k_c\kuroten k_{c^\prime})_I \right] }$ by
$\displaystyle{\prod^N_{a=3} D(k_{am})}$,
 (  see eq. (\ref{eq:damping2})), we find that the amplitude $A_{N}$ contains
 an overall multiplicative factor
\begin{equation}
 D  \left( \sum_{a=3}^{N} k_{am} \right) = 
 \exp\left[-\frac{\pi}{2}\sum_{I}\left|\beta_{I}\right|
  \left((  \sum_{a=3}^{N} k_{a})\kuroten
 (\sum_{b=3}^{N} k_{b})\right)_{I}\right]\;\;,
\label{eq:damping}
\end{equation} 
 which depends upon the total photon momentum alone.
We have thus seen that the string amplitude in fact has resummed
 and lifted the approximate infinite degeneracy of the spectrum by evaluating
  its effect as an exponential factor
 and that this lifting has rendered
the net g.d.f. of the amplitude to depend only upon the total momentum of the
incoming photons.

  We turn to the four point amplitude in the zero slope limit. 
After lifting the infinite degeneracy due to the lowest bosonic mode, we still
 have the contributions from several nearly degenerate states due
 to the lowest fermionic modes. 
  See eq. (5.8) of \cite{CIMM2} 
 for the complete formula in the zero slope limit
  which contains the above  contributions as well.
Picking up only those parts of the amplitude in which
the state with the lowest mass (tachyon) participates, we find
\begin{eqnarray}
\lim   {\cal A}_{4}&=& -2i (2\pi)^{p+1}
\prod_{\mu=0}^{p}\delta \left(\sum_{a=1}^{4}k_{a\mu}\right)
D \left( k_{3m} + k_{4m} \right) \;
   \exp\left(\frac{i}{2}\theta^{\mu\nu}k_{1\mu}k_{2\nu}
              +\frac{i}{2}\theta^{MN}k_{3M}k_{4N}\right) 
\nonumber\\
 && \left[ \frac{1}{t-m_{T}^{2}}
\frac{1}{2}\left\{\left(k_{2}-(k_{1}+k_{4})\right) \pdot \zeta_{3}
  {}-ik_{3}\ppdot J\zeta_{3} \right\} \right.\nonumber\\
&& \hspace{6em} \left\{\left((k_{2}+k_{3})-k_{1}\right) \pdot
     \zeta_{4} -ik_{4}\ppdot J\zeta_{4} \right\} \nonumber\\
&&\hspace{1em}  +\frac{1}{s}\left\{
   \left(
   (k_{2}-k_{1})\pdot\zeta_{3}-i(k_{3}+k_{4})\ppdot J\zeta_{3}
  \right) k_{3}\prdot\zeta_{4}\right.\nonumber\\
&&\hspace{4em}-\left(
       (k_{2}-k_{1})\pdot\zeta_{4}
     {}-i(k_{3}+k_{4})\ppdot J\zeta_{4} \right)
     k_{4}\prdot\zeta_{3}  \nonumber\\
&&\hspace{4em}
   + \left( \frac{1}{2}(k_{3}-k_{4})\pdot (k_{1}-k_{2})
    {}-ik_{3}\ppdot J k_{4}
    \right)    \zeta_{3}\prdot\zeta_{4}\bigg\} \bigg]
   \nonumber\\
&&  +\left(k_{3}\leftrightarrow k_{4};
   \zeta_{3}\leftrightarrow\zeta_{4}\right)
   \;\;.
\label{eq:st4ampstring}
\end{eqnarray}

 Let us give an explicit connection  between the g.d.f. and
 the noncommutative projector soliton, which is a key
 observation to our work.
The g.d.f. is rewritten as 
\begin{eqnarray}
 D(k_{m})    &=&\exp  \left( -\frac{1}{4}
\sum^{\frac{p^\prime}{2}}_{I=\frac{p+2}{2}}
 \mid \theta^{2I-1,2I}  \mid (k_{2I-1}k_{2I-1}+k_{2I}k_{2I}) 
 \right) \nonumber\\
&=&\prod^{\frac{p^\prime}{2}}_{I=\frac{p+2}{2}}
\tilde{\phi_0}(k_{2I-1},k_{2I} ;\theta^{2I-1,2I}) \;\;.
\end{eqnarray}
Observe that
\begin{eqnarray}
\label{eq:observe}
  2 \pi \mid \theta \mid
\tilde{\phi_0}&=&\int d^2x e^{ik_1x^1+ik_2x^2}
\phi_0(x^1,x^2 ;\theta) \;\;,  \nonumber \\
\phi_0(x^1,x^2 ;\theta)&=&
2e^{-\frac{1}{ \mid \theta \mid}((x^1)^2+(x^2)^2)} \;\;.
\end{eqnarray} 
Function $\phi_0(x^1,x^2 ;\theta)$ is the projector soliton solution of the
 noncommutative scalar field theory discussed in \cite{GMS}. 
It satisfies $\phi_0\ast \phi_0=\phi_0$ and  is
represented as a ground state projector $|0\rangle\langle0|$ in the 
Fock space representation of noncommutative algebra $[x^1, x^2]=i\theta$.
In \cite{GMS}, $\phi_0$ is discussed as a soliton solution of noncommutative 
scalar field theory in the large $\theta$ limit.  In our discussion,
 Fourier transform of $\phi_0$  is seen to appear for all values of $\theta$.

  Eq. (\ref{eq:lima3}) tells us  that 
the g.d.f.  $D(k_{3m})$ is a form factor of the $Dp$-brane
  of size $\sqrt{\mid \beta_{I} \mid}$ by noncommutative $U(1)$ current,
 which can be written as
\beq
  \left(  \Phi^\dagger   \stackrel{\leftrightarrow}{\partial}^\mu \Phi ,
-i\partial_{n} \left(\Phi^\dagger J^{mn}\Phi \right)  \right)
   \;\;,
\label{eq:ncu1c}
\eeq
  using the scalar field $\Phi(x^{\mu},x^{m})$ discussed in the next section.
Putting together this fact and the obsevation  of the last paragraph,
 we identify the $Dp$-brane
 in the zero slope limit with the noncommutative soliton.
See \cite{HKLM} for this identification from string field theory.

\section{Dp brane and the Projector Soliton of
 Noncommutative Scalar Field Theory}
We now give a field theoretic derivation of the properties
 of the string amplitude
in the zero slope limit given by eqs. (\ref{eq:lima3}), (\ref{eq:damping}) and 
(\ref{eq:st4ampstring}). We will show that an adequate description is given in 
perturbation theory of low energy effective 
action (LEEA) proposed in \cite{CIMM2} by specifying proper initial and final 
states associated with the scalar field $\Phi(x^\mu, x^m)$.

In \cite{CIMM2}, the following action has been proposed:
\begin{eqnarray}
&&S = S_0 + S_1~,\nonumber\\
&&\mbox{with}\quad
S_0 = \frac{1}{g_{YM}^{\ 2}}\int d^{p'+1}x \sqrt{-G}
\left\{ -\left(D_{\mu}\Phi \right)^{\dag}\ast\left(D^{\mu}\Phi\right)
       {}-m^2 \Phi^{\dag} \ast \Phi
       {} -\frac{1}{4} F_{MN} \ast F^{MN}\right\}~,\nonumber\\
&&\hspace{3em}
S_1 = \frac{1}{2g^{\ 2}_{YM}}\int d^{p'+1}x \sqrt{-G}
\Phi^{\dag} \ast F_{mn} J^{mn} \ast \Phi~,
\label{eq:leea}
\end{eqnarray}
where 
\begin{eqnarray}
&&D_{\mu}\Phi =\partial_{\mu}\Phi-iA_{\mu} \ast \Phi~,
\qquad
\left( D_{\mu} \Phi \right)^{\dag}
  =\partial_{\mu}\Phi^{\dag}+i \Phi^{\dag}\ast A_{\mu}~,\nonumber \\
&&F_{MN}=\partial_{M}A_{N}-\partial_{N}A_{M}
    {}-i \left[A_{M},A_{N}\right]_{\ast}~,\quad
 \left[A_{M}, A_{N}\right]_{\ast}= A_{M} \ast A_{N}
    {}-A_{N} \ast A_{M} \;\;.
\end{eqnarray}
  Here $A_M(x^\mu, x^m)$ is a $(p^\prime+1)$-dimensional vector
 field which corresponds to 
noncommutative $U(1)$ photon and $\Phi(x^\mu,x^m)$ is a 
 scalar field which corresponds
to the ground state tachyon of the $p-p^\prime$ open string with 
$\displaystyle{m^2=   - \lim_{\alpha^\prime\to 0}
   (1-  \displaystyle{ \sum_I \nu_I }) / 2\ap } $. 
Reflecting the fact that the tachyon momenta are constrained
 to lie in $p+1$ dimensions, 
the Lorentz index of the kinetic term for the scalar field runs from 0
 to p
 and there
is no kinetic term for the remaining $p^\prime-p$
 directions.  From now on, we set
  $g_{YM}$  to $1$.

It is elementary to compute the three point tree amplitude 
 from ${\cal L}_{\rm int}(\Phi, A_M)$ :
\bea
 {\cal A}_3 &=&
  i \int d^{(p^\prime-p)}K_m^{(f)}
    \int d^{(p^\prime+1)}x^M \sqrt{-G} \,
    {}_{\rm  sol}\langle \! \langle - K^{(f)}_m |
      \otimes {_{\rm tach}\langle}-k_{2\mu} |  \nonumber\\
  &&  \left\{ \frac{1}{2}\Phi^\dagger
      \ast {_{\rm vec}\langle}\,0\,|   F_{mn}J^{mn}
                         |k_{3M}\rangle_{\rm vec}
      \ast \Phi   \right. \nonumber\\
  && \left. -i\Phi^\dagger  \left( \ast {_{\rm   vec} \langle}\,0\,|
     A_\mu |k_{3M}\rangle_{\rm vec}\ast  \stackrel{\rightarrow}{\partial}^\mu
 -\stackrel{\leftarrow}{\partial}^\mu 
\ast {_{\rm   vec} \langle}\,0\,| A_\mu |k_{3M}\rangle_{\rm vec}\ast  \right)
 \Phi  \right\}  |k_{1\mu}  \rangle_{\rm tach}
\otimes | \, 0 \, \rangle \! \rangle_{\rm  sol} \nonumber\\
 &=& -i \left( \frac{1}{\sqrt{-G}} \right)^{2}
     (2\pi)^{p+1}\delta^{(p+1)} \left( \sum^3_{a=1}k_{a\mu} \right) 
     \exp\left(\frac{i}{2}\theta^{\mu\nu}k_{1\mu} k_{2\nu}\right)
       u^\ast(k_{3m})u(0)   \\
  &&  \prod_{a=1,2}\frac{1}{ \sqrt{(2\pi)^{p}2\omega_{{\vec k}_{a}}} }
      \frac{1}{\sqrt{(2\pi)^{p} 2|{\vec k}_{3}|}}
  \left( (k_{2}-k_{1})\prdot \zeta_{3}-ik_{3}\ppdot J\zeta_{3}  \right)
 \nonumber  \;\;.
\label{eq:a3leea}
\eea
Here the Fock space associated with the vector and
 those with the $x^{\mu}$ and $x^{m}$
 dependent part of  the scalar field $\Phi(x^{\mu},x^{m})$
 are designated by $vec, tach, sol$ respectively.
Eq.(\ref{eq:lima3}) from string theory and eq.(\ref{eq:a3leea})  computed from 
 eq.(\ref{eq:leea}) agree completely provided 
\beq
\label{eq:id}
u^\ast(k_m)u(0)= D(k_m) \;\;,  \;\;\;{\rm or} \;\;  u(k_m)= D(k_m)
\eeq
  The  momentum space wave function in soliton sector is identified with
  the g.d.f. and hence is equal to Fourier image of the distribution of
  the noncommutative soliton.

Let us next see that the N-point tree amplitude
 obtained from this field
 theory contains the g.d.f. whose argument is the total momentum. 
Carrying out the Wick contractions and using the propagator 
 which contains the delta function, 
 we find that the field theory $N$ point amplitude contains
the following factor residing in the soliton sector: 
\beqn
   \int d^{(p^\prime-p)}K^{(f)}_m 
\prod^L_{a=3} \left( \int d^{(p^\prime-p)}x^{m}_a \right)
{}_{\rm sol}\langle\langle -K^{(f)}_m  \mid \phi^\dagger(x^{m}_3)\ast 
e^{iq_{3}\ppdot x_{3}}
\ast \delta^{(p^\prime-p)}(x_{3}-x_{4}) \nonumber \\
  \ast  e^{ iq_{4} \ppdot  x_{4}} \ast \cdots 
 \ast \delta^{(p^\prime-p)}(x_{L-1}-x_{L})\ast e^{ iq_{L}\ppdot x_{L}}
 \ast \phi(x^{m}_L)|K^{(i)}_m=0 \rangle \rangle_{\rm sol} \;\;\;. 
\eeqn
  Thanks to the delta function propagator, this equals
\beqn
   &=& \int d^{(p^\prime-p)}K^{(f)}_m \int d^{(p^\prime-p)}x^{m}
   \exp  \left( \frac{i}{2} \displaystyle{ 
\sum_{ {\scriptstyle a,b=3} \atop{\scriptstyle a < b} }^{L} 
\sum_{m, n=  p+1}^{p^{\prime}}
  \theta^{mn} q_{a m} q_{b n} }   \right)   \nonumber\\
   & &{_{\rm sol}\langle\langle}-K^{(f)}_m|\phi^\dagger(x^{m})
  \ast   \exp  \left( {i  \left( {\displaystyle \sum^L_{a=3}}q_{a}
 \right) \ppdot x } 
 \right)  \ast  \phi(x^m)|
  K^{(i)}_m=0\rangle\rangle_{\rm sol} \nonumber \\
 &=& \int d^{(p^\prime-p)}K^{(f)}_m   \delta^{(p^\prime-p)} \left( 
K^{(f)}_m  + (\sum^L_{a=3}q_{a}) \right)
 u^\ast ( - K^{(f)}_m)u(0)
  \exp  \left( \frac{i}{2} \displaystyle{ \sum_{   {\scriptstyle a,b=3} 
 \atop{\scriptstyle a < b} }^{L}
 \sum_{m, n= p+1 }^{p^{\prime}}
  \theta^{mn} q_{am} q_{bn} }     \right)
 \nonumber \\
 &=& D \left(\sum^L_{a=3}k_{am} \right) \exp  \left(  \frac{i}{2}
  \displaystyle{  \sum_{  {\scriptstyle a,b=3}  
 \atop{\scriptstyle a < b}}^{L}
 \sum_{m, n=p+1}^{p^{\prime}}
  \theta^{mn} q_{am} q_{bn} }  \right)    \;\;\;.
\eeqn 

Finally, let us check that the tree four point amplitude (the pole part) in 
fact agrees with string answer.
After the Wick contraction and the position space integration,
we find
\begin{eqnarray}
 {\cal A}_{4} \!\!\! &=&\!\!\!
   (2\pi)^{p+1} \delta^{p+1}\left(\sum_{a=1}^{4}k_{a\mu}\right)
     \, \exp\left(\frac{i}{2}\theta^{\mu\nu}k_{1\mu}k_{2\nu}\right)
     u^{\ast}(k_{3M}+k_{4M}) \, u(0)
 \nonumber\\
 && \left(\frac{1}{\sqrt{-G}} \right)^{3} 
   \prod_{a=1,2}
    \frac{1}{\sqrt{ (2\pi)^{p}2\omega_{\vec{k}_{a}} } }
     \prod_{b=3,4}
     \frac{1}{\sqrt{ (2\pi)^{p^{\prime}} |\vec{k}_{b}|} }
       \left( {\bf a}^{(t,u)}_{4} + {\bf a}^{(s)}_{4}   \right)~,
\end{eqnarray}
where
\begin{eqnarray}
{\bf a}_{4}^{(t,u)}\!\!\!&=&\!\!\!
     \frac{-i}{t-m^{2}}
    \left\{ (k_{2}-(k_{1}+k_{4}))\pdot \zeta_{3}-ik_{3}\ppdot J\zeta_{3}
    \right\} \nonumber\\
  && \hspace{4em}
    \left\{(k_{2}+k_{3})-k_{1}) \pdot \zeta_{4}
      {} -ik_{4} \ppdot J\zeta_{4} \right\} \,
      \exp\left(\frac{i}{2}\theta^{MN}k_{3M}k_{4N}\right)
     \nonumber \\
  &&+\left(k_{3}\leftrightarrow k_{4};
           \zeta_{3}\leftrightarrow \zeta_{4} \right)~, \\
{\bf a}_{4}^{(s)} \!\!\! &=& \!\!\!
    \frac{-i}{s} \,\,
    2\left[ \left( (k_{2}-k_{1})\pdot \zeta_{3}
           {}-i(k_{3}+k_{4})\ppdot J\zeta_{3}\right)
                k_{3}\prdot \zeta_{4} \right. \nonumber\\
  && \hspace{3em} - \left( (k_{2}-k_{1}) \pdot \zeta_{4}
           {}-i(k_{3}+k_{4}) \ppdot J \zeta_{4} \right)
                k_{4} \prdot \zeta_{3} \nonumber \\
  && \hspace{3em}\left.
       + \left( \frac{1}{2} (k_{3}-k_{4}) \pdot (k_{1}-k_{2})
           {} -ik_{3} \ppdot J k_{4} \right) 
           \zeta_{3} \prdot \zeta_{4} \right]
           \exp\left( \frac{i}{2}\theta^{MN}k_{3M}k_{4N}\right)
           \nonumber\\
   && + \left( k_{3} \leftrightarrow k_{4};
               \zeta_{3} \leftrightarrow \zeta_{4} \right)~.
\end{eqnarray}
  This expression agrees with   eq.(\ref{eq:st4ampstring}).
 
 We have thus shown that perturbation theories of two different kinds in fact
  agree.


\end{document}